\documentstyle[12pt,epsf]{article}

\let\th=\theta

\newcommand{\beq}{\begin{equation}}
\newcommand{\eeq}{\end{equation}}
\newcommand{\be}{\begin{equation}}
\newcommand{\ee}{\end{equation}}
\newcommand{\bea}{\begin{eqnarray}}
\newcommand{\eea}{\end{eqnarray}}

\newcommand{\nbox}{{\,\lower0.9pt\vbox{\hrule \hbox{\vrule height 0.2 cm \hskip
0.2 cm \vrule height 0.2 cm}\hrule}\,}}

\DeclareFixedFont{\xiiss}{OT1}{cmss}{m}{n}{12}
\DeclareFixedFont{\ixss}{OT1}{cmss}{m}{n}{9}
\DeclareFixedFont{\cmrnine}{OT1}{cmr}{m}{n}{9}

\newcommand{\CC}{\hbox{\xiiss C\kern-.4emI}}
\newcommand{\RR}{\hbox{\xiiss R\kern-.45emI}}
\newcommand{\ZZ}{\hbox{\xiiss Z\kern-.4emZ}}
\newcommand{\CCs}{\hbox{\ixss C\kern-.4emI}}
\newcommand{\ZZs}{\hbox{\ixss Z\kern-.4emZ}}
\newcommand{\pa}{\partial}

\newcommand{\pasl}{\pa\kern-.55em /}
\newcommand{\da}{{$\overline{D2}~$}}
\newcommand{\dd}{{$D2-\overline{D2}~$}}
\def\href#1#2{#2}
\begin{document}
\begin{titlepage}
\title{
\vskip-3cm
        \begin{flushright}
        \begin{small}
        RU-NHETC-2000-35\\
         EFI-2000-37\\
        hep-th/0010016\\
        \end{small}
        \end{flushright}
        \vspace{1.cm}
Tachyon Condensation in Noncommutative Gauge Theory \\
}

\author{
Per Kraus\thanks{e-mail: \tt pkraus@theory.uchicago.edu}\\
        \small\it Enrico Fermi Institute,
\small\it University of Chicago,
\small\it Chicago, IL 60637
\\
\\
Arvind Rajaraman\thanks{e-mail: \tt arvindra@physics.rutgers.edu} \\
        \small\it Department of Physics and Astronomy,
\small\it Rutgers University,
\small\it Piscataway, NJ 08855
\\
\\
Stephen Shenker\thanks{e-mail: \tt shenker@het1.stanford.edu}\\
        \small\it Department of Physics,
\small\it Stanford University,
\small\it Stanford, CA 94305.
}

\maketitle
\begin{abstract}
\noindent
We show that the decay of the $D2$-\da system with
 large worldvolume magnetic fields can be described in
noncommutative gauge theory.  Tachyon condensation in this
system describes the annihilation of $D2$-\da into
$D0$-branes.  From the $2+1 $ dimensional point of view,
this is the decay of a nonabelian magnetic flux into
vortices.
The semiclassical approximation
is valid over a long period of the decay.
Our analysis  allows us to clarify earlier results in the
literature related to tachyon condensation and
noncommutative gauge theory.  
\end{abstract}

\end{titlepage}

\section{Introduction}
\baselineskip 0.65cm

Studying the condensation of tachyons on unstable D-brane systems
has led to an improved understanding of the configuration space of string
theory.  For instance, it has been established that D-branes can be
described as solitons made of open strings
\cite{Sen:1998ii,Witten:1998cd,Horava:1999jy,Harvey:2000qu,Harvey:2000tv,
deMelloKoch:2000ie,Moeller:2000jy,deMelloKoch:2000xf,Moeller:2000hy},
 and that the closed string
vacuum can be constructed as a nonperturbative state in open string field
theory \cite{Kostelecky:1990nt,Sen:2000nx,Berkovits:2000hf,Moeller:2000xv,
DeSmet:2000dp,Rastelli:2000iu,Kostelecky:2000hz}.

However, many questions remain regarding the physics around the stable
vacuum.  These include the nature of the spectrum of open string field
theory expanded around this vacuum
\cite{Sen:2000hx,Taylor:2000ku,Iqbal:2000qg}, the emergence of closed strings
\cite{Yi:1999hd,Bergman:2000xf,Harvey:2000jt},
and the potential role of strong coupling phenomena 
\cite{Yi:1999hd,Bergman:2000xf}.
Addressing these
questions in open string field theory is challenging because of the infinite
number of component fields which acquire expectation values\footnote{This
problem was solved recently \cite{Gerasimov:2000zp,Kutasov:2000qp}.}, and the
infinite number of higher derivative terms in the action which describe
dynamics.

Recently, it was shown that considering tachyon condensation in the presence
of large B-fields --- or equivalently large worldvolume magnetic fields ---
 can lead to great simplifications
\cite{Harvey:2000jt,Dasgupta:2000ft,Witten:2000nz}.  In particular,
techniques of noncommutative geometry
\cite{Connes:1998cr,Seiberg:1999vs,Gopakumar:2000zd}
allow an exact construction of D-branes
as open string solitons, reproducing the correct tensions and fluctuation
spectra \cite{Harvey:2000jt}.
In the present paper we will similarly show that the presence of
large magnetic fields simplifies the description of the annihilation of
D-branes via tachyon condensation.  We will consider the $D2$-\da system
in IIA string theory, with large but opposite sign magnetic fields on
each of the branes.  This system is unstable and will decay into $D0$-branes.
The simplifying feature of the large magnetic field is that the decay can
be described entirely within noncommutative Yang-Mills theory.

The Yang-Mills description of our system is provided by Matrix Theory
\cite{Banks:1997vh}.
The relation between Matrix Theory, noncommutative field theory, and
tachyon condensation was noted in \cite{Gopakumar:2000zd,Harvey:2000jt}
and discussed in more detail in \cite{Seiberg:2000zk}.
Also, the decay of the $D\overline{D}$ system in Matrix Theory was studied in
\cite{Awata:1999sy,Massar:2000jp},
before the recent developments in noncommutative field theory.

Some of the questions concerning tachyon condensation in open string field
theory involve understanding what happens to the open string degrees
of freedom on the original unstable D-brane, what the correct variables
are to describe the ground state of the system, and whether the ground
state is unique
\cite{David:2000uv,Sochichiu:2000rm,Gopakumar:2000rw,Sen:2000vc,Sen:2000tg}.
These questions become most sharp in the case of decay
to the closed string vacuum.  In the present setup we will be left with
$D0$-branes after the decay, and it is clear that the original $2+1$
dimensional noncommutative Yang-Mills degrees of freedom turn into
the $0+1$ dimensional degrees of freedom describing the $D0$-branes.
Furthermore, modulo gauge transformations, the classical ground states are
uniquely
labelled by the positions of the $D0$-branes.

The rest of this paper is organized as follows.  In section \ref{string}
we discuss the $D2$-\da system in the limit of large worldvolume magnetic
fields.  The point is that in this limit the energy released by the
decay of the system becomes small, so that rather than having to shift
the entire string field we can concentrate on a small subset of modes.
We  then work out the spectrum of light open strings in this system, for
 later comparison with the noncommutative field theory.
Matrix Theory provides a description of the $D2$-brane in term of
$D0$-branes, as we review in section \ref{matrix}.  As emphasized recently
in \cite{Seiberg:2000zk},  the $0+1$ dimensional Matrix Theory action for
this system
becomes, when rewritten using the Weyl correspondence, a $2+1$ dimensional
noncommutative $U(1)$ gauge theory.  Conversely, one can rewrite
the $D2$-brane field theory as an action for a collection of $D0$-branes.
It is this fact which allows us to describe the decay of the  $D2$-\da system
into $D0$-branes in terms of noncommutative gauge theory, since it shows that
both the initial and final states can be described as particular
configurations in the same underlying action. We then confirm that the
fluctuation spectrum of the gauge theory precisely matches the open
string spectrum computed from free string theory.  In section \ref{decay}
we discuss of the decay process itself.  From a $2+1$ dimensional
point of view, this corresponds to the decay of a constant magnetic
field into a gas of vortices, as pointed out in \cite{Awata:1999sy}.
Describing the
actual decay process explicitly is difficult, so we instead exhibit
some paths in configuration space interpolating between the initial
and final states.  We also discuss an approach involving level truncation
in the noncommutative gauge theory.   Finally, in section \ref{conclude}
we conclude with some discussion.

Field theory models of tachyon condensation have been considered
 in \cite{Zwiebach:2000dk}.  The fact that $D\overline{D}$
annihilation in the presence of fluxes can be described within field
theory was discussed in \cite{Taylor:2000ek}.  The noncommutative field
theory description of tachyon condensation was studied recently in the 
$D2-D0$ system in \cite{agms}, and in the $D5-\overline{D5}$ system in 
\cite{Tatar}.

\section{The \dd system in string theory}
\label{string}

We are  interested in describing the \dd system as a bound
state of $D0$-branes.  A $D2$-brane constructed in this way has on its
worldvolume a magnetic field whose strength is proportional to the
density of $D0$-branes, as follows from the coupling
\be
S^{D2}_{wz} = -i \mu_2 \int \! d^3x \, C^{(1)} \wedge 2 \pi \alpha' (F^++B),
\ee
where the $+$ superscript on $F$ will distinguish the field strength on the
$D2$ from that on the \da.
Given the gauge invariant combination $F+B$ we can make an arbitrary split
into  an $F$ part and a $B$ part, and we will find it convenient to label
the background as $B_{12}=$ constant,  $F=0$, though we of course allow
$F$ to vary dynamically around this background.
 We  describe the background field in terms of the dimensionless
parameter $b$,
\be
b = 2\pi\alpha' B_{12}.
\ee
We have chosen $x^1,x^2$ as the spatial worldvolume directions.
The requirement that the density of $D0$ charge be positive implies $b>0$.

The \da has an opposite sign for this coupling,
\be
S^{\overline{D2}}_{wz} = +i \mu_2 \int \! d^3x \, C^{(1)} \wedge 2 \pi
\alpha' (F^-+B).
\ee
One way to understand the sign change is to recall that a \da can be obtained
from a $D2$ by a $\pi$ rotation in a $x^i-x^a$ plane, where $x^i$ is
a spatial wordvolume direction and $x^a$ is a transverse direction.  The
implication is that in order to describe the \da as a bound state of
$D0$-branes rather than $\overline{D0}$-branes we should take $F^-_{12}
+B_{12}<0$
on the  \da.   We accomplish this by taking a background value of $F^-_{12}$
such that
\be
\label{dbf}
F^-_{12} +B_{12} = -b.
\ee
An important point is that our system thus does not correspond simply to a
$D2$-\da in a background NS B-field ---  this would induce negative
$D0$ charge on the \da.  It was necessary to turn on the background worldvolume
magnetic field in order to remedy this.

Thinking of the $D2$ as a bound state of $D0$'s, it is useful to work
out the ``binding'' energy, in particular in the regime $b \gg 1$.
This is given by the  Born-Infeld term for the $D2$, which is
\be
S_{BI} = - \mu_2 \int \! d^3x \, \sqrt{-\det [g_{ab}+2\pi\alpha'
(F^++B)_{ab}]} .
\ee
Inserting the background field $b$ and expanding for $b \gg 1$ gives
\be
\label{BIexpand}
S_{BI} = - \mu_2 b \int \! d^3x \, \sqrt{g}
  - {\mu_2 \over 2 b} \int \! d^3x \, \sqrt{g}.
\ee
The first term represents the energy of an equivalent density of $D0$-branes,
and the second term is the excess. For our purposes, the notable feature
 is that the latter quantity becomes small (as a density) for $b \gg 1$.
Actually, it is more appropriate to express the binding energy in terms of
the open string coupling and metric of \cite{Seiberg:1999vs},
$G_s \sim b g_s$ and $\sqrt{G} \sim b^2 \sqrt{g}$.   Then, taking
$b \rightarrow \infty$ while keeping the open string quantities fixed one
finds that the binding energy density  scales as $1/b^2$.  So in this
regime, the decay of the unstable \dd system releases a very small energy
density. This means that we can expect to be able
to describe this process in the zero slope limit, that is by a noncommutative
gauge theory description.  We will indeed find that this is the case.

\subsection{Open string spectrum}

We now work out the spectrum of open strings on the coincident \dd system,
focussing on modes which will survive in the zero slope limit.
A similar analysis appears in \cite{Seiberg:1999vs}.
We again
take the worldvolume to lie in the $X^1, X^2$ directions, and denote
transverse directions by $X^i$.  It is convenient to use the complex
coordinate $Z = (X^1 +i X^2)/\sqrt{2}$.   As discussed above, we take
\be
(F^+ + B)_{12} = b, \quad (F^- + B)_{12} = -b.
\ee
The spectrum of $22$ and $\bar{2}\bar{2}$ strings is unaffected by $b$.  We now
consider the $2\bar{2}$ strings.
The boundary conditions on worldsheet fields are
\bea
\sigma &=0: \quad\quad \partial_\sigma Z -ib \partial_\tau Z = 0, \quad
 \partial_\tau X^i = 0. \cr
\sigma &=\pi: \quad\quad \partial_\sigma Z +ib \partial_\tau Z = 0, \quad
 \partial_\tau X^i = 0.
\eea
The solution is
\bea
Z &=& i \sum_{n=-\infty}^{n=\infty} \left[
e^{i(n+\nu)(\sigma-\tau)} + e^{i\nu \pi}e^{-i(n+\nu)(\sigma +\tau)}\right]
{\alpha_{n+\nu} \over n+\nu}~+c~,  \cr
&& \cr
X^i &=& i\sum_{n=-\infty}^{n=\infty}\left[
e^{in(\sigma-\tau)} - e^{-in(\sigma +\tau)}\right]
{\alpha_n^i \over n}~,
\eea
where
\be
e^{i\nu \pi} = {1 + ib \over 1- ib}.
\ee
As $b$ is taken from $0$ to $\infty$, $\nu$ ranges from $0$ to $1$.  For
$b \gg 1$,
\be
\nu \approx 1 - {2 \over \pi b}.
\ee
The open string spectrum is determined from
\be
\alpha' m^2 = {\rm (oscillator~ energy)} + a.
\ee
In the Ramond sector the ground state energy $a$ vanishes as usual,
$a_R =0$.   In the Neveu-Schwarz sector we find
\be
a_{NS} = -{1 \over 4} - {1 \over 2}\left|\nu-{1 \over 2}\right|,
\ee
which for large $b$ behaves as
\be
b \gg 1: \quad\quad a_{NS} \approx -{1 \over 2} + {1 \over  \pi b}.
\ee

An important point is that the bosonic creation operator $\alpha_{-(1-\nu)}$
raises the energy by $2/(\pi b)$ for large $b$.  So given any light state
we can produce a  tower of light states by applying $\alpha_{-(1-\nu)}$
an arbitrary number of times.

Now we work out the spectrum, concentrating on states with $\alpha' m^2
\sim 1/b$ or less, since these will survive the zero slope limit.
We work in light cone gauge, taking $X^1, \cdots, X^8$ as transverse
directions.

We begin with the R sector. The ground
state consists of two massless, opposite chirality, $4$ component SO(6)
spinors.
One of these is allowed by the GSO projection.  The other is projected out,
but we get an allowed state by acting with a worldsheet fermion.  To obtain
a light state we can act with $\psi_{-(1-\nu)}$.  Now, given these two light
states we act with $\alpha_{-(1-\nu)}$ to generate the following
towers of states
\bea
4~~ {\rm of}~~~~~  \alpha' m^2 &=& {2n \over \pi b}~, \cr
&& \cr
4~~ {\rm of}~~~~~  \alpha' m^2 &=&
{2 \over \pi b}+ {2n \over \pi b}, \quad n = 0,1,2...~.
\eea

Next we consider the NS sector.  For large $b$ the ground state has energy
$a_{NS} \approx -{1 \over 2} + {1 \over  \pi b}.$   First, we need to know
whether this state is allowed by the GSO projection.  When quantizing
strings stretching between $D2$ and \da one takes the opposite GSO projection
compared to that for $D2$-$D2$ strings.  There is another consideration,
which is that when we take $b$ from 0 to $b \gg 1$ the two fermionic modes
$\psi_{\pm (1/2 - \nu)}$ cross, shifting the fermion number of the ground
state by $1$.  Together, this implies that for $b \gg 1$ the NS ground
state is GSO projected out.   Allowed light states are generated
by acting on the ground state with
\be
 \psi^\dagger_{\nu -1/2},~~\psi_{-(3/2 - \nu)},~~ {\rm and}~~
\psi^i_{-1/2}.
\ee
Acting on these with $\alpha_{-(1-\nu)}$ we generate the towers
\bea
1 ~~ {\rm of}~~~~~ \alpha' m^2 &=& -{1 \over \pi b} + {2n \over \pi b}~, \cr
&& \cr
1 ~~ {\rm of}~~~~~ \alpha' m^2 &=& {3 \over \pi b} + {2n \over \pi b}~, \cr
&& \cr
6 ~~ {\rm of}~~~~~ \alpha' m^2 &=& {1 \over \pi b} + {2n \over \pi b}~.
\eea

\vspace{.2cm}
Along with each state above, we should include
 its partner coming from
strings beginning on \da and ending on $D2$.
Note that there is then a single complex tachyon in the spectrum,
signalling the
instability of the $D2$-\da system.
All states in the theory besides those just worked out
 have $\alpha' m^2 \sim O(1)$ for large $b$.

The zero slope limit described in \cite{Seiberg:1999vs} consists of taking
\be
\alpha' \sim  \epsilon^{1 \over 2} \rightarrow 0, \quad\quad
{1 \over b} \sim  \epsilon^{1 \over 2} \rightarrow 0,
\ee
with $g_{\mu\nu}$ fixed.\footnote{This is a coordinate transformed version
of the limit presented on p. 12 of \cite{Seiberg:1999vs}.}  In this limit,
the light states we have computed remain in the spectrum while all other
states are removed.
We will see in the next section that the field theory describing the light
states
is a $2+1$ dimensional noncommutative gauge theory with a background magnetic
flux.

\section{Matrix Theory description of $D2$-\da}
\label{matrix}

In this section we study the description of the $D2$-\da system as an unstable
bound state of D0-branes.  The construction we use is identical to that used
in Matrix Theory \cite{Banks:1997vh}, although our interpretation is
somewhat different.  We will
not consider the strong coupling quantum effects which lead to an eleven
dimensional large $N$ limit.  Instead our considerations will be purely
classical, as is appropriate in the $g_s \rightarrow 0$ limit at the
fixed length scale $b^{-{1 \over 2}}$.  We will have more to say about
the validity of the classical approximation in section \ref{decay}.

\subsection{Review of $D2$-branes}

We begin by reviewing the construction of $D2$-branes.  The action
describing a collection of $N$ D0-branes is, in units with $2\pi \alpha'=1$,
\be
\label{D0action}
S = {T_0 \over 2} \int \! dt\, {\rm Tr}\left\{
D_0 X^I D_0 X^I +{1 \over 2}[X^I,X^J]^2 - {i \over 2}\lambda D_0 \lambda
+{1 \over 2}\lambda \Gamma^0 \Gamma^I [X^I,\lambda]\right\}
\ee
with $I,J = 1, \cdots, 9$; $T_0 = {\sqrt{2 \pi} \over g}$ is the mass of a
$D0$-brane; $D_0 = \partial_t -i [A_0, \cdot]$; and $X^I$, $\lambda$
are Hermitian $N \times N$ matrices.   We henceforth take the $N
\rightarrow \infty$ limit, so that the matrices become operators on
an infinite dimensional Hilbert space.  Without loss of generality,
it is convenient to take this Hilbert to be that of a one dimensional
harmonic oscillator with a basis of energy eigenstates $|n\rangle$.

The background describing a flat, infinite $D2$-brane in the $X^1,X^2$
plane is given by $X^1 =x^1,~ X^2=x^2$,  with
\be
\label{eqd2}
[x^1,x^2] = i\theta.
\ee
It is sometimes convenient to write the solution in terms of
\be
z = { x^1 +ix^2 \over \sqrt{2}} = \sqrt{\theta} a,
\ee
with $a$ the annihilation operator, $[a,a^\dagger]=1$.
To make contact with noncommutative field theory, consider a general
bosonic fluctuation around this background.
\be
A_0, \quad X^a = x^a + \theta \epsilon^{ab}A_b, \quad X^i,
\ee
$a,b = 1,2$, and $i=3, \cdots, 9$.  Expanding out the action (\ref{D0action}),
keeping all bosonic terms, gives \cite{Banks:1997nn,Keski-Vakkuri:1998ec}
\be
\label{expand}
S = T_0 \int \! dt \, {\rm Tr}\left\{
-{1 \over 4}G^{\mu\alpha}G^{\nu \beta}(F+\Phi)_{\mu\nu}(F+\Phi)_{\alpha\beta}
+{1 \over 2}G^{\mu\nu}D_\mu X^i D_\nu X^j \right\}~.
\ee
In this action Greek indices run over $0,1,2$.  We have defined the following
quantities:
\bea
G_{\mu\nu} &=& {\rm diag}(1,-\theta^{-2},-\theta^{-2})~, \cr
F_{0a} &=& \partial_t A_a - i \theta^{-1} \epsilon_{ab}[x^b,A_0]
+i [A_0,A_a]~, \cr
F_{12} &=& i\theta^{-1}[x^a,A_a] + i[A_1,A_2]~, \cr
D_a X^i &=& i \theta^{-1} \epsilon_{ab}[x^b,X^i] - i[A_a,X^i]~, \cr
\Phi_{12} &=& -\theta^{-1}~.
\eea
The $0+1$ dimensional action (\ref{expand}) can be rewritten as a
$2+1$ dimensional field theory by using the Weyl correspondence
(see, {\it e.g.} \cite{Gopakumar:2000zd}). Under this correspondence
operators on Hilbert space are replaced by functions on
$I\!\!R^2$ according to the rules
\bea
AB &\rightarrow & \left. (A \star B)(x) \equiv e^{{\theta \over 2}\epsilon^{ab}
\partial_a \partial_{b'}} A(x)B(x')\right|_{x'=x} \cr
{\rm Tr} &\rightarrow & {1 \over 2 \pi \theta}\int \! d^2x.
\eea
A useful formula following from the definition of the $\star$ product
is
\be
x^a \star f - f \star x^a = i \theta \epsilon^{ab}\partial_b f
\ee
for any function $f$.
Using these rules,
the action takes the form (\ref{expand}) but with the replacements
\baselineskip 0.80cm
\bea
T_0 &\rightarrow& {T_0 \over 2 \pi \theta}~, \cr
dt~ {\rm Tr} &\rightarrow& d^3x~, \cr
F_{\mu\nu} &\rightarrow&
\partial_\mu A_\nu - \partial_\nu A_\mu + i[A_\mu,A_\nu]~, \cr
D_\mu X^i &\rightarrow& \partial_\mu X^i - i [A_\mu,X^i]~,
\eea \baselineskip 0.65cm
and with all products replaced by $\star$ products.

The resulting action is just the noncommutative field theory describing
the zero slope limit of a D2-brane in a
background field $B_{12} = \theta^{-1}$, with the particular choice of
$\Phi$ parameter (explained in \cite{Seiberg:1999vs}) given by $\Phi = -B$.
This shows quite explicitly that a configuration of $D0$-branes satisfying
(\ref{eqd2}) represents a $D2$-brane in a background $B$ field. It
also demonstrates background independence \cite{Seiberg:2000zk},
in that we get the same action
whether we represent the configuration in terms of $D0$-brane quantum
mechanics, or in terms of $D2$-brane $2+1$ dimensional field theory.  The
two descriptions correspond to expanding the same underlying actions around
two different backgrounds.  It is this property of background independence
that will allow us to describe the decay of the $D2$-\da system in terms of a
$2+1$ dimensional field theory, since it guarantees that the final state
after the decay --- $D0$-branes --- are describable in terms of these
degrees of freedom.

\subsection{The $D2$-\da system}

A \da in Matrix Theory is given by satisfying (\ref{eqd2}) with an additional
minus sign,
\be
[X^1,X^2] = -i\theta,
\ee
corresponding to reversing the orientation of the brane.  The combined
$D2$-\da system then corresponds to a direct sum of the individual solutions,
\be
\label{dd}
X^1=\left(
\begin{array}{cc}
x^1 & 0 \\
0  &  x^1
\end{array} \right), \quad\quad
X^2=\left(
\begin{array}{cc}
x^2 & 0 \\
0  &  -x^2
\end{array} \right).
\ee
Background independence implies that we can think of the  $D2$-\da system
as being a particular flux configuration in U(2) $2+1$ dimensional
noncommutative field theory.  This is easily accomplished by writing
(\ref{dd}) as
\be
\label{newrep}
X^a = x^a \otimes {\bf 1}_{2\times 2} + \theta \epsilon^{ab}A_b,
\ee
with $A_a$ a U(2) gauge field with background value
\be
\label{backb}
A_1 = { 2 x^2 \over \theta}\left(
\begin{array}{cc}
0 & 0 \\
0 & 1
\end{array} \right).
\ee
Thus the decay of the $D2$-\da system is equivalent to the decay of a
particular nonabelian magnetic flux in the noncommutative field theory.
The background magnetic flux breaks U(2) down to U(1) $\times$ U(1), which
is indeed the correct gauge group for $D2$-\da.  As we will see shortly,
the instability of the system is manifested by a tachyonic off diagonal
mode.

Similarly to what we saw in the previous subsection, expanding the action
(\ref{D0action}) around the background (\ref{dd}) gives a field theory action
of the form (\ref{expand}), but now with all fields taking vales in the
Lie algebra of $U(2)$, and with the background magnetic flux (\ref{backb}).

The background gauge field (\ref{backb}) corresponds to a magnetic field
on the \da of strength,
\be
F^-_{12} = - {2 \over \theta}.
\ee
This can be compared to our discussion in section 2 once we translate the
present noncommutative field strength used in this this section to the
commutative field strength used there.  Using the relation
 \cite{Seiberg:1999vs},
\be
F^{{\rm com}}_{12} = {F^{nc}_{12} \over 1 + \theta F^{nc}_{12}},
\ee
and $\theta^{-1}=b$ we find that the commutative field strength
satisfies the relation (\ref{dbf}) up to an overall minus sign.
The reason for the apparent sign discrepancy is that turning on the
noncommutative gauge field (\ref{backb}) necessarily reverses the orientation
of the worldvolume coordinates on the \da, as explained in
\cite{Cornalba:1999ah}.  Once this is taken into account (\ref{dbf}) is
satisfied.
Recall that this relation enforced the
condition that the \da be composed of the same positive density of
$D0$-branes as is the $D2$-brane.

\subsection{Fluctuations around the $D2$-\da system}

In section \ref{string} we worked out the spectrum of light open string modes
in the  $D2$-\da system, and we noted that the spectrum should be
reproduced in the zero slope limit.   Checking this is a matter of
computing the quadratic fluctuation spectrum of the action
(\ref{D0action}) around the background (\ref{dd}).  This computation
has already been done in the Matrix Theory literature
\cite{Aharony:1997bh,Lifschytz:1997rw}.   To
compare, use the relations $b = c = 2\pi z^2$ and $2 \pi \alpha' =1$.
(In comparing,  one also needs to be careful about factors of two
involving complex boson versus real boson masses).
The spectra match exactly, as expected.

It is useful to identify some of the fluctuations explicitly.  The
eigenvectors of the fluctuation matrices are given in
\cite{Aharony:1997bh}.  Writing the background (\ref{dd}) in complex
coordinates,    the tachyon corresponds to the fluctuation
\be
Z = {X^1 + iX^2 \over \sqrt{2}} = \sqrt{\theta} \left(
\begin{array}{cc}
a & 0 \\
 T P_0 & a^\dagger
\end{array}
\right),
\ee
with $[a,a^\dagger]=1$, and $P_0$ being the projection operator
onto the harmonic oscillator ground state, $P_0 = |0\rangle\langle0|$.
The action
for $T$ is found to be
\be
\label{tachact}
S = {T_0 \theta \over 2} \int \! dt \, \left\{|\dot{T}|^2
+2 \theta |T|^2 - {1 \over 2} \theta |T|^4 \right\},
\ee
corresponding to a complex tachyon of mass $m^2 = -2 \theta$,
(or $\alpha' m^2 = -{1 \over \pi b}$ in the notation of section
\ref{string}).


\section{Decay to $D0$-branes}
\label{decay}

The $D2$-\da system is unstable and will decay.  By charge
conservation, it must decay to a collection of $D0$-branes.  The
decay will be initiated by the condensation of the tachyon
identified in the previous section.  However, a complicating
feature of the system we are considering is that tachyon
condensation will not occur homogeneously over the $D2$-\da worldvolume,
as is seen from the following considerations.  The tachyon has
charge $(1,-1)$ under the unbroken $U(1) \times U(1)$ gauge group on the
$D2$-\da.  As we have discussed, building this system up out of
$D0$-branes means that there are background magnetic fields in each
$U(1)$, with a positive sign in  the $U(1)$ associated to the
$D2$, and a negative sign in  the $U(1)$ associated to the \da.
In terms of the relative $U(1)$, the tachyon is a charged scalar
field in a background magnetic field, obeying an equation of the schematic form
\be
(\partial_\mu - iA_\mu)^2\,T = V'(T).
\ee
The spatial dependence of the background $A_\mu$ shows that the
tachyon will condense inhomogeneously.

Physically, what happens is clear: the tachyon will condense into
a collection of vortices, with the original constant magnetic field
becoming localized in the cores.  On the other hand, a vortex on the
 $D2$-\da system is nothing other than a $D0$-brane \cite{Sen:1998ii}.
So this process represents the $D2$-\da system decaying into $D0$-branes.

While the final vortex configuration sounds rather complicated when
expressed in the $2+1$ dimensional field theory language,
 it is simple when we go
back to  $D0$-brane quantum mechanics.
%
%
If we wish to find a static configuration to which the system will
decay, we can simply minimize the energy
\bea
H=-{T_0 \over 2}\  {\rm Tr}\  {1 \over 2}[X^I,X^J]^2,
\eea
which has the obvious solution $[X^I,X^J]=0$.
We therefore expect that the  $D2$-\da system
will evolve to a configuration where the matrices commute. This is
a system of $D0$-branes.
In the language of the BFSS Matrix Theory, we have a description
of the annihilation of a membrane-antimembrane system into
gravitons.

\subsection{Validity of the classical approximation}

Our analysis of the decay will be carried out entirely at the classical
level.  This is justified in the limit of large $b$ and small $g_s$ as
follows.  First of all, before the $D2$-\da has begun to decay we can make the
system arbitrarily weakly coupled at the magnetic length scale
$\th^{{1 \over 2}}$ by taking $g_s$ sufficiently small.  We now show that the
final $D0$-brane system behaves classically for a long time.  The final
state consists of $D0$-branes with spacing $\Delta x \sim b^{-{1 \over 2}}
\sim \th^{{1 \over 2}} $
moving with characteristic velocity $v$.   $v$ is determined by noting from
(\ref{BIexpand})  that the energy per $D0$-brane released by the decay
is $E \sim \mu b^{-2} \sim \mu \th^2$, where $\mu$ is the $D0$-brane mass.
This implies $v \sim \th$.  Now, note that the time for the $D0$-branes
to collide is $t \sim \Delta x/v \sim \th^{-{1 \over 2}}$ which is large
in our limit.  Furthermore, the Compton wavelength of the $D0$-branes,
$\lambda \sim (\mu \th)^{-1}$, can be taken to be much smaller than
the inter-brane spacing $\Delta x$ by taking $g_s$ small.  So in this
limit there is a long time scale over which it is valid to think of the
$D0$-branes as slowly moving classical particles.
Large $N$ effects do not destroy this semiclassical picture because of the
large size ($\sim N^{1/2}$ ) of the planar array of $D0$-branes.
We should also note
that
the decay of the system
occurs on a much shorter time
 scale, $t \sim \sqrt{\th}$, as seen, for example, from
(\ref{tachact}).   So for a long time after the decay has occured we can
treat the $D0$-branes classically; eventually though, they collide, become
quantum mechanically entangled, and form bound states.  We will have
nothing more to say about this final stage in the evolution of our system,
as it is outside the realm of the weak coupling approximation.
Another way of describing this semiclassical decoupled regime is to specify
that the three relevant
length scales, the string length $l_s$,  the magnetic length $l_m =
\sqrt{\theta}$, and the eleven dimensional planck length $l_{11} = g_s^{1/3}
l_s$ must obey the relation $l_{11}<<l_m<<l_s$.

Energetically, the final $D0$-branes can be distributed arbitrarily
in the $x^1, x^2$ plane.  However, given that we start with a
homogeneous $D2$-\da system it is clear that the  distribution
resulting from an actual decay will also be nearly homogeneous.
But since such a distribution is not naturally written in terms of
the diagonal matrices $X^1$ and $X^2$ we will focus on more special
distributions.  Our point in the following is just to exhibit some
paths in configuration space along which the original $D2$-\da system can
be deformed into a zero energy state.


To analyse the decay, we shall have to consider
intermediate matrix configurations where the
branes have partially decayed. This is difficult
to do if we are working in the limit where the
matrices are infinite dimensional. It will be
useful, therefore, to also consider finite dimensional
representations of branes, and later take the
size of the brane     to infinity.

The problem is that it is not possible to solve
the equation (\ref{eqd2}) with finite matrices.
We can however find solutions of  a related problem
as in \cite{Banks:1997vh,hoppe,mathguy}.
If we have matrices satisfying $[A_2,A_1]={2\pi i\over N}$, then
the matrices $U=e^{iA_1},V=e^{iA_2}$, satisfy
\bea
\label{UVeqn}
UV=\omega VU
\eea
where $\omega = e^{{2\pi i\over N}}$.

Now although we cannot find finite dimensional representations
of (\ref{eqd2}), we can find finite dimensional
representations of (\ref{UVeqn}). These are the clock and shift
operators of rank N
\bea
U_{i,i+1}&=&1, \quad~ i=1,2\cdots N-1
\nonumber
\\
U_{N,1}&=&1
\nonumber
\\
V_{i,i}&=&\omega^{i-1}
\eea
and the rest of the elements are zero.

We can therefore define our finite dimensional approximation
to (\ref{eqd2}) by the matrices $x_1^N,x_2^N$ satisfying
\bea
x_1^N=-i \ ln(V),\quad~ x_2^N=-i \ ln(U).
\eea

\subsection{Representation 1}
One natural choice of vacuum
state is
\bea
\label{rep1}
X^1=0,\quad~ X^2=0
\eea
corresponding to all of the $D0$-branes being at the origin.
More generally, we can take $X^1$ and $X^2$ to be any diagonal
matrices.

One can write a sequence of classical matrix configurations
 connecting the initial $D2$-\da  state  with the final state
of diagonal matrices. We will consider an
initial
configuration with the $D2$-\da represented by the
rank N approximation.

\bea
X^1=\left(\begin{array}{cc}x_1^N&0\cr
0&x_1^N
\end{array}
\right),
\quad\quad
X^2=\left(\begin{array}{cc}x_2^N&0\cr
0&-x_2^N
\end{array}
\right)
~~~~~~~~~~~~~~
\nonumber
\\
\Downarrow~~~~~~~~~~~~~~~~~~~~~~~~~~~~~~~~~~~~~~~
\nonumber
\\
\Downarrow~~~~~~~~~~~~~~~~~~~~~~~~~~~~~~~~~~~~~~~
\nonumber
\\
~~~~~~~~~~~~~~~~~~~~~~~~~~~~~~~~~~
\nonumber
\\
X^1=\left(\begin{array}{cccc}x_1^{(N-k)}&0&0&0\cr
0&D_1^{(k)}&0&0\cr
0&0&x_1^{(N-k)}&0\cr
0&0&0&D_2^{(k)}
\end{array}
\right),
~~~~~~~~~~~~~~~~~~~~~~~~~~~~~~~~~~
\nonumber
\\
~~~~~~~~~~~~~~~~~~~~~~~~~~~~~~~~~ \nonumber
\\
X^2=\left(\begin{array}{cccc}
x_2^{(N-k)}&0&0&0\cr
0&D_3^{(k)}&0&0\cr
0&0&x_2^{(N-k)}&0\cr
0&0&0&D_4^{(k)}
\end{array}
\right)
\nonumber
\\ \nonumber
\\
\Downarrow~~~~~~~~~~~~~~~~~~~~~~~~~~~~~~~~~~~~~~~
\nonumber
\\
\Downarrow~~~~~~~~~~~~~~~~~~~~~~~~~~~~~~~~~~~~~~~
\nonumber
\\
X^1=\left(\begin{array}{cc}D_1^N&0\cr
0&D_2^N
\end{array}
\right),
\quad\quad
X^2=\left(\begin{array}{cc}D_3^N&0\cr
0&D_4^N
\end{array}
\right)
~~~~~~~~~~~~~~
\nonumber
\eea

where the $D_i$ are diagonal matrices.

\subsection{Representation 2}
Another representation of the vacuum using commuting matrices is
\bea
\label{rep2}
X^1=\left(\begin{array}{cc}x_1&x_2\cr
x_2&x_1
\end{array}
\right),
\quad\quad
X^2=\left(\begin{array}{cc}x_2&x_1\cr
x_1&x_2
\end{array}
\right).
\eea

There is a decay process similar to
the one described above
\bea
X^1=\left(\begin{array}{cc}x_1^N&0\cr
0&x_1^N
\end{array}
\right),
\quad\quad
X^2=\left(\begin{array}{cc}x_2^N&0\cr
0&-x_2^N
\end{array}
\right)
~~~~~~~~~~~~~~
\nonumber
\\
\Downarrow~~~~~~~~~~~~~~~~~~~~~~~~~~~~~~~~~~~~~~~
\nonumber
\\
\Downarrow~~~~~~~~~~~~~~~~~~~~~~~~~~~~~~~~~~~~~~~
\nonumber
\\
X^1=\left(\begin{array}{cccc}x_1^{(N-k)}&0&0&0\cr
0&x_1^{(k)}&0&x_2^{(k)}\cr
0&0&x_1^{(N-k)}&0\cr
0&x_2^{(k)}&0&x_1^{(k)}
\end{array}
\right)
~~~~~~~~~~~~~~~~~~~~~~~~~~~~~~~~~~
\nonumber
\\
\nonumber
\\
~~~~~~~~~~~~~~~~~~~~~~~~~~~~~~~~~
X^2=\left(\begin{array}{cccc}
x_2^{(N-k)}&0&0&0\cr
0&x_2^{(k)}&0&x_1^{(k)}\cr
0&0&x_2^{(N-k)}&0\cr
0&x_1^{(k)}&0&x_2^{(k)}
\end{array}
\right)\nonumber
\\
\Downarrow~~~~~~~~~~~~~~~~~~~~~~~~~~~~~~~~~~~~~~~
\nonumber
\\
\Downarrow~~~~~~~~~~~~~~~~~~~~~~~~~~~~~~~~~~~~~~~
\nonumber
\\
X^1=\left(\begin{array}{cc}x_1^N&x_2^N\cr
x_2^N&x_1^N
\end{array}
\right),
\quad\quad
X^2=\left(\begin{array}{cc}x_2^N&x_1^N\cr
x_1^N&x_2^N
\end{array}
\right).
~~~~~~~~~~~~~~
\nonumber
\eea

\section{The 2d version of the decay}

\subsection{A puzzle}
To make contact with the analysis of the
previous sections, we need to take
the limit of $N\rightarrow \infty$.

In this limit the decay process in
the first representation looks like
the process
\bea
X^1=\left(\begin{array}{cc}x_1&0\cr
0&x_1
\end{array}
\right),
\quad\quad
X^2=\left(\begin{array}{cc}x_2&0\cr
0&-x_2
\end{array}
\right)~~~~~~~~~~~~
\nonumber
\\
\Downarrow~~~~~~~~~~~~~~~~~~~~~~~~~~~~~~~~~~~~~~~
\nonumber
\\
\Downarrow~~~~~~~~~~~~~~~~~~~~~~~~~~~~~~~~~~~~~~~
\nonumber
\\
X^1=\left(\begin{array}{cccc}x_1&0&0&0\cr
0&D_1^{(k)}&0&0\cr
0&0&x_1&0\cr
0&0&0&D_2^{(k)}
\end{array}
\right),
\quad\quad
X^2=\left(\begin{array}{cccc}
x_2&0&0&0\cr
0&D_3^{(k)}&0&0\cr
0&0&x_2&0\cr
0&0&0&D_4^{(k)}
\end{array}
\right). \nonumber
\eea

Therefore we start from a $D2$-\da system
and produce $D0$-branes; in the 2d language, this
is the production of vortices on the branes.  These
vortices are composed purely of gauge field, and exist
only in the noncommutative field theory.  See \cite{agms}
for a detailed discussion.

This leads us to an apparent puzzle. It would seem that the
process of creating vortices on branes should require
positive energy. In fact, if we consider the formation
of a vortex on a single $D2$-brane, it
was explicitly shown that this requires a positive
energy. This is as it should be, since we know
that a $D2$-brane on its own is stable.

We can belabour the point further by calculating how
much the energy increase in Matrix theory. The
energy of a membrane in the lightcone Hamiltonian is
$M^2/N$, where N is the momentum in the lightcone direction.
If we remove a zero-brane, the energy becomes
$M^2/(N-1)$. Hence the energy increases by
a factor ${N\over N-1}$.

To resolve this puzzle, let us consider the formation
of a single vortex on a single $D2$-brane  which has been regulated as we
have been doing so far. Then the first guess could be
that the process is  described
by the matrix configurations
\bea
X^1=x_1^N,\quad\quad X^2=x_2^N
~~~~~~~~~~~ \nonumber
\\
\Downarrow~~~~~~~~~~~~~~~~~~~~~~~~
\nonumber
\\
\Downarrow~~~~~~~~~~~~~~~~~~~~~~~~
\nonumber
\\
\nonumber
\\
X^1=\left(\begin{array}{cc}x_1^{(N-1)}&0\cr
0&0
\end{array}
\right),
\quad\quad
X^2=\left(\begin{array}{cc}x_2^{(N-1)}&0\cr
0&0
\end{array}
\right)
\eea
where the total rank of both matrices is N.

However, we must now also include that the fact that
membrane charge is conserved. In other
words, we want $Tr[X^1,X^2]$ to be conserved. If we
use the matrix configurations above, then the two
matrices in the limit $N\rightarrow \infty$ do not
have the same trace. The trace in the second matrix is
${N-1\over N}$ times the trace of the first.

To remedy this, we should consider the process
\bea
X^1=x_1^N,\quad\quad  X^2=x_2^N \quad\quad\quad\quad\quad
~~~~~~~~~~~
\\
\Downarrow~~~~~~~~~~~~~~~~~~~~~~~~\quad\quad\quad\quad\quad
\nonumber
\\
\Downarrow~~~~~~~~~~~~~~~~~~~~~~~~\quad\quad\quad\quad\quad
\nonumber
\\
X^1=\sqrt{{N\over N-1}}\left(\begin{array}{cc}x_1^{(N-1)}&0\cr
0&0
\end{array}
\right),
\quad\quad
X^2=\sqrt{{N\over N-1}}\left(\begin{array}{cc}x_2^{(N-1)}&0\cr
0&0
\end{array}
\right).
\eea

Now comparing the energies, we find that the ratio of energies
of the configurations
is
\bea
{Tr[X^1,X^2]^2_{final}\over Tr[X^1,X^2]^2_{initial}}={N\over N-1}.
\eea
Hence the energy indeed increases by a factor  ${N\over N-1}$, as
expected.

Now in the $D2$-\da system, the total membrane
charge is zero. Hence the process that we wrote
earlier is consistent without any extra factors.
In this case, the energy of forming a vortex is indeed
negative. In fact, we can calculate the ratio
of energies of the intermediate configuration
to the initial one to be
\bea
{Tr[X^1,X^2]^2_{final}\over Tr[X^1,X^2]^2_{initial}}={N-k\over N}.
\eea
Hence the energy decreases, and the decay is allowed.

\subsection{ Decay in representation 1}
Having resolved this puzzle, let us turn to the
interpretation of the decay in representation 1.

The final state representing the vacuum has $X^1=X^2=0$.
>From the 2d point of view,
we can think of this as a
fluctuation about two D2-branes
(\ref{newrep}),
\be
X^a = x^a \otimes {\bf 1}_{2\times 2} + \theta \epsilon^{ab}A_b = 0.
\ee
Hence
\bea
A_1 = {1 \over \theta} x^2  \otimes {\bf 1}_{2\times 2}, \quad\quad
A_2 = -{1 \over \theta} x^1  \otimes {\bf 1}_{2\times 2}.
\eea

We therefore see that in this vacuum, only the gauge fields
are nonzero, and they are precisely chosen so that
the covariant derivative acting on any field
vanishes (as $[X^i, \cdot]=0$).  What has happened is that the original
magnetic field (\ref{backb}), living purely in the second $U(1)$, has
redistributed itself symmetrically in the two $U(1)$s via the formation of
vortices.  As pointed out in \cite{Gopakumar:2000rw} this vanishing covariant
derivative indicates that the open string degrees of freedom can no longer
propagate and have become unphysical.

Furthermore, this vacuum has a classical $U(\infty)$ symmetry.
In the case of decay to nothing a vacuum state with this symmetry was
proposed in \cite{Gopakumar:2000rw}.  By analogy, we will refer to the
vacuum in representation $1$ as the ``GMS vacuum''.

We note, though,  that more generally the vacuum in this representation
corresponds to $X^a$ diagonal, not zero.  In such a vacuum $U(\infty)$ is
spontaneously broken (and in fact a gauge transformation takes this vacuum to
the one discussed below).  The covariant derivative is not zero,  but now
allows independent motion of each $D0$-brane. Two dimensional propagation has
ceased.

Of course after very long times strong quantum fluctuations become important
and the $U(\infty)$ symmetry is restored,  but these two phenomena are not
linked.

As an aside let us remind the reader that the unbroken $U(\infty)$ state of
Matrix theory and the spontaneously broken $U(\infty)$ state of an arbitrarily
shaped matrix membrane (where $U(\infty)$ is interpreted as the group of area
preserving diffeomorphisms) provide an instructive analog of Witten's ideas
about the existence an unbroken (perhaps topological) phase of gravity,  as
well as a more conventional spontaneously broken phase.  These ideas may well
be important in the search for background independent formulations of quantum
gravity.

\subsection{ Decay in representation 2}
The second representation of the decay has a
final state where the off diagonal matrix is
nontrivial. This implies that the tachyon is
nontrivial in this vacuum. We would like to
relate this vacuum to the sort of vacuum proposed by Sen, in which
the tachyon condenses homogeneously while the gauge fields remain at zero.
However, this is difficult to do in this context, as we have
already shown that there is not expected to be a solution
with a constant tachyon.

The solution instead represents, not a constant
tachyon field, but rather, a configuration of tachyon vortices.
These tachyon vortices each carry one unit of $D0$-brane charge,
and hence this corresponds to a distribution
of $D0$-branes.

It is helpful to diagonalize (\ref{rep2}) in order
to specify the $D0$-brane distribution.
Working in a diagonal basis for $x_1$, $x_2$ is represented as
$x_2 = -i \partial_{x_1}$.  The eigenvectors of $X^1$ and $X^2$ are
\be  \vec{v}_1 =
\left(\begin{array}{c} f_\lambda(x_1) \cr f_\lambda(x_1) \end{array} \right),
\quad\quad
\vec{v}_2 =
\left(\begin{array}{c} \overline{f_\lambda(x_1)} \cr
-\overline{f_\lambda(x_1)} \end{array} \right),
\ee
with eigenvalues
\bea
X^1 \vec{v}_1 &=& \lambda \vec{v}_1,  \quad\quad
X^2 \vec{v}_1 = \lambda \vec{v}_1, \cr
X^1 \vec{v}_2 &=& \lambda \vec{v}_2,  \quad\quad
X^2 \vec{v}_2 = -\lambda \vec{v}_2,
\eea
where we have defined
\be
f_\lambda(x_1) = e^{-{i \over 2}(x_1 -\lambda)^2}.
\ee
Normalizability requires $\lambda$ to be real.  The distribution
of $D0$-branes thus corresponds to two diagonal lines
 crossing at the origin of the $X^1, X^2$ plane.

\subsection{Relation between representations}

We have provided two representations of the final vacuum
state, and representations of the decay
into these vacuum states. In the first representation only the gauge
field condensed, while in the second representation the tachyon condensed
as well.
We now would like
to understand the relation between these
two representations.

In the quantum mechanical version of the decay,
it is straightforward to see that these two
representations are gauge equivalent. Since
any pair of commuting matrices can be diagonalized, one
can diagonalize the matrices (\ref{rep2}) and
obtain
a diagonal matrix as in (\ref{rep1}).

In the 2d language it is hard to see this
gauge equivalence, because the $U(2)$ symmetry
of the two $D2$-brane system is spontaneously
broken in the $D2$-\da system. This is a
further indication that the  ``right''
variables to describe
noncommutative theories are the matrix model variables, as pointed
out by Seiberg
\cite{Seiberg:2000zk}.

Secondly, since we are identifying representation 1 with the GMS
description of the vacuum, and representation 2 with the
vacuum
described by Sen, we would
like to argue that these two are actually the same,
i.e. the GMS vacuum is gauge equivalent to the Sen
vacuum.

This is not a precise statement, since we have not
identified the exact vacuum corresponding to the
Sen vacuum in our framework. However, we have shown
that diagonal matrices are gauge equivalent to the other
configurations where the tachyon is nontrivial.
Furthermore, all such configurations, which have
an off diagonal component turned on, can be diagonalized
to a form where only the gauge field is turned on.
Hence the Sen vacuum, which must have a nonzero profile
for the tachyon field, is gauge equivalent to a configuration which
has a zero tachyon field, but a nontrivial gauge field.

This configuration of gauge fields is classically describable
as a distribution of vortices over the plane.
The Sen vacuum is expected to be gauge equivalent
to a homogeneous distribution of
vortices. As such, it would appear to be
different from the GMS vacuum, where all the vortices
are at one point. These vacua are connected by
a marginal deformation, where we move the zero-branes
from one configuration to the other.

Hence we can more
accurately describe the GMS and Sen vacua as being on
the same moduli space of vacua.

Although we have not discussed the quantum mechanics of
the system so far, it is easy to see from general
considerations that there will be solutions corresponding to
threshold bound states in the final solution.

\subsection{Level truncation}

By turning on large magnetic fields on the worldvolumes of the
$D2$ and \da and taking the zero slope limit, we have vastly reduced
the number of degrees of freedom of the system.  However, the problem
is still sufficiently complicated that it is difficult to explicitly
track the evolution of the system to its ground state.  To make the
problem even simpler we can employ level truncation, as is done in
open string field theory \cite{Kostelecky:1990nt,Sen:2000nx}.
This corresponds to solving the equations of
motion of some number of light modes, and setting all heavier modes to
zero.  In our case the simplest level truncation consists of retaining
only the tachyon, with the action being given by (\ref{tachact}).
Minimizing the tachyon potential gives,
\be
|T| = \sqrt{2}.
\ee

We now argue that this configuration corresponds, roughly, to a
$D2$-\da with a hole around the origin.  First, in operator language
the tachyon configuration
is proportional to the projection operator $P_0$;
in terms of functions this corresponds to a Gaussian centered at the origin
\cite{Gopakumar:2000zd}.
To get an idea of whether the $D2$-\da has decayed in the region where
the tachyon has condensed, we will introduce an extra $D0$-brane into
the system to act as a probe.  The point is that before $T$ condensation
there is a tachyonic mode coming from strings connecting the auxilliary
$D0$-brane to the $D2$-\da system, but we will see that this tachyon
acquires a positive $m^2$ once $T$ condenses, indicating that the
$D2$-\da system has decayed to $D0$-branes near the origin.
Thus we consider in complex coordinates
\be
Z = \sqrt{\theta} \left( \begin{array}{ccc}
0 & t & 0 \\
0 & a & 0 \\
0 & T P_0 & a^\dagger \end{array} \right).
\ee
$t$, the $D0-D2$
tachyon, is a row vector with a nonzero entry in
its first component.   The potential for $t$ is
\be
V(t) = 4 \theta^2 (|T|^2 - 1 )|t|^2.
\ee
This gives the desired result: $t$ is tachyonic   at $T=0$, and acquires
  positive $m^2$ at $|T| = \sqrt{2}$ .

It would be interesting to include more modes in the level truncation.
For instance at the next level we should include the $U(1) \times U(1)$
gauge fields.
We expect that this will describe a $D2$-\da with an expanding hole
that eventually fills the worldvolume once all modes are retained.


\section{Comments}
\label{conclude}
We conclude with a few comments:
\begin{enumerate}

\item In the case of tachyon condensation into ``nothing'' there is the
question
of what happens to the open string degrees of freedom on the original D-brane.
Expanding around the stable minimum, the spectrum must either consist of states
with nonperturbatively heavy masses, or states that can be interpreted as
closed
strings, but how either scenario is realized is not well understood.  In
the present
case the system decays into $D0$-branes, and it is clear that the original
degrees of freedom are repackaged as strings ending on these $D0$-branes.  The
effect of tachyon condensation is to convert degrees of freedom which look
$2+1$ dimensional into ones which look $0+1$ dimensional.   It has been
proposed
\cite{Gopakumar:2000rw}
that in the case of decay to nothing the final configuration looks $0+0$
dimensional.
Studying the decay to nothing seems to require dealing with the full open
string
field theory, rather than just the zero slope limit as was done here.
In addition the scenario in \cite{Gopakumar:2000rw} requires an understanding
of the dynamics that produces large amounts of flux.  In the current work the
$D0$-branes are put in "by hand."

\item  We have studied the decay of the $D\overline{D}$ system,
but there are other
unstable D-branes in type II string theory: $D(2p+1)$-branes in IIA, and
$D(2p)$-branes in IIB.  It is natural to wonder whether these branes can
also be constructed out of lower dimensional branes, {\it e.g.} $D0$-branes
in IIA.
This seems to be complicated for the following reason.  $D0$-brane charge shows
up on the $D2$-brane as magnetic flux, and a similar story holds for the other
$D(2p)$-branes.  However on an unstable $D1$-brane, for example, $D0$-charge
arises via spatial variation of the tachyon, $S_{wz}= \int C^{(1)}\wedge dT$.
Constant $D0$ charge density then requires a linearly varying $T$, but this is
not a solution to the equations of motion.


\end{enumerate}

\vspace{0.2in}

\paragraph{Acknowledgments: }
P.K. is supported by NSF grant PHY-9901194,  A.R. is supported in part by DOE
grant DE-FG02-96ER40559, S.S. is supported in part by NSF grant PHY-9870115. We
thank Rajesh Gopakumar, Matthew Kleban, 
Emil Martinec, John McGreevy, Shiraz Minwalla, Rob Myers, and Andy
Strominger  for discussions, and the Aspen Center for Physics for hospitality
during the initial stages of this work.

\end{document}